# Impact of Unintentional Oxygen Doping on Organic Photodetectors

Julie Euvrard [a], Amélie Revaux [a,*], Alexandra Cantarano [a], Stéphanie Jacob [a], Antoine Kahn [b], and Dominique Vuillaume [c]

[a] Univ. Grenoble Alpes, CEA-LITEN, Grenoble, 38000, France

[b] Dept. of Electrical Engineering, Princeton University, Princeton, NJ, 08544, USA

[c] IEMN, CNRS, Univ. Lille, Villeneuve d'Ascq, 59652, France

Oxygen plasma is a widely used treatment to change the surface properties of organic layers. This treatment is particularly interesting to enable the deposition from solution of poly(3,4-ethylenedioxythiophene)-poly(styrenesulfonate) (PEDOT:PSS) on top of the active layer of organic solar cells or photodetectors. However, oxygen is known to be detrimental to organic devices, as the active layer is very sensitive to oxygen and photo-oxidation. In this study, we aim to determine the impact of oxygen plasma surface treatment on the performance of organic photodetectors (OPD). We show a significant reduction of the sensitivity as well as a change in the shape of the external quantum efficiency (EQE) of the device. Using hole density and conductivity measurements, we demonstrate the p-doping of the active layer induced by oxygen plasma. Admittance spectroscopy shows the formation of trap states approximately 350 meV above the highest occupied molecular orbital of the active organic semiconductor layer. Numerical simulations are carried out to understand the impact of p-doping and traps on the electrical characteristics and performance of the OPDs.

1. Introduction

Organic photodetectors (OPDs) have achieved characteristics that compete with those of amorphous silicon devices, and can be partially or fully processed using solution printing techniques on transparent and flexible substrates [1–3]. The major challenges organic devices still face include stability and processing issues, which need to be solved to target large-scale industrialization. The first organic photodetectors and solar cells were developed according to a standard structure [4], whereby electrons and holes are collected at the top and bottom electrodes, respectively. The electron transport layer (ETL) is typically made of a low work-function material, which is sensitive to oxygen and moisture [5,6]. The hole transport layer (HTL) is usually made of poly(3,4-ethylenedioxythiophene)-poly(styrenesulfonate) (PEDOT:PSS), which is known to degrade the underlying indium tin oxide (ITO) layer due to its acidic nature [7]. To overcome those issues, the structure has been inverted and PEDOT:PSS is deposited on top of the active layer [8].

---

* Corresponding author. Tel.: +33 438784593.
 E-mail address: amelie.revaux@cea.fr (A. Revaux).

The main challenge of the inverted structure is the HTL processing. Water-based PEDOT:PSS presents the advantage of solution processing, but is incompatible with the hydrophobic properties of most organic active layers. To allow PEDOT:PSS deposition from solution on organic layers, additives need to be used (e.g. Surfynol [9]) or surface treatments need to be carried out on the active layer to change its surface tension. Among those treatments, $O_2$ plasma is the most used technique [10,2], as it improves the active layer wettability by creating functional groups at the surface [11].

However, it is widely reported that oxygen is detrimental to the performances of organic devices because of photo-oxidation processes [12–15]. Oxygen is known to p-dope organic semiconductors, increasing their conductivity [16] but also introducing gap states [17–20]. Oxygen-related degradations are currently one of the major issues in ageing studies of organic photodetectors and solar cells [12]. It is then crucial to understand the impact of oxygen on the device initial performances when used as surface treatment on top of the active layer.

The group of Baierl et al. [21] compared an inverted structure processed with $O_2$ plasma treatment with a standard structure processed without treatment. In their study, they show that $O_2$ plasma does not lead to performance deterioration of organic photodiodes, explaining the reduction of the EQE at 500 nm by the different transmission of gold and ITO. However, the EQE spectra were not fully analyzed and the current density in the direct regime could not be directly compared due to different injection barriers. In the present work, we process the same inverted structure with and without $O_2$ plasma treatment, enabling direct evaluation of plasma impact. We combine electrical characterization with simulations to determine and understand the influence of the $O_2$ plasma.

2. **Experimental methodology**

Two devices are processed, one without and one with $O_2$ plasma treatment. The diodes are processed on ITO-coated glass substrates. Polyethylenimide (PEIE) is used to reduce the ITO work function. The active layer consists of poly[(4,8-bis-(2-ethylhexyloxy)-benzo(1,2-b:4,5-b')dithiophene)-2,6-diyl-alt-(4-(2-ethylhexanoyl)-thieno[3,4-b]thiophene-)-2-6-diyl)] (PBDTTT-c) and phenyl-C61-butyric acid methyl ester (C60-PCBM) as donor and acceptor materials, respectively. This structure is detailed and characterized elsewhere [22]. Depending on the structure studied, an $O_2$ plasma treatment precedes the PEDOT:PSS deposition. The surface treatment is carried out for 60 s with a power of 500 W, a pressure of 1.5 bar and a flow of 200 sccm. The plasma exposure time is chosen to offer the best wettability of the active layer [23]. The samples are covered with a metallic grid during the plasma treatment. The PEDOT:PSS Orgacon HIL 1005 purchased from Agfa is deposited by lamination following the process described by Gupta et al. [10]. PEDOT:PSS is spin-coated on an $O_2$ plasma treated PDMS stamp. The plasma treatment is required to change the surface tension of the stamp from hydrophobic to hydrophilic. The PDMS stamp covered with PEDOT:PSS is then left in air for 20 min drying. After bringing the PEDOT:PSS and active layer into contact with the active layer, the full stack is annealed at 85°C for 5 min to reverse the PDMS surface to hydrophobic. The PDMS stamp is carefully removed after cooling down the substrate to ambient temperature. The devices are encapsulated with glass in the glove box using an epoxy glue.

The TLM devices are processed on gold-coated poly(ethylene 2,6- naphthalate) (PEN). The 30 nm thick gold electrodes are patterned using photolithography. An $O_2$ plasma treatment is used to clean the surface before spin-coating the blend layer. The same solution and deposition technique are used for the OPD and this structure. The appropriate surface treatment is then carried out on each layer and the devices are encapsulated in a glove box.

Current-voltage and capacitance-voltage measurements are carried out using a Keithley 2636A source meter and an Agilent E4980A LCR meter respectively. The set-up used to measure the EQE spectra of OPDs is described elsewhere [24]. Admittance spectroscopy measurements are performed in a vacuum chamber with a chuck cooled down to 77 K with liquid nitrogen. A heater is used to reach the required temperature. An impedance analyzer Keysight E4990A is used for the admittance measurements. For this study, C(f) and G(f) characteristics are carried out from 20 Hz to 10 MHz with an AC signal amplitude of 10 mV and a bias of 0 V.

3. **Results and Discussion**

    3.1. **Impact of $O_2$ plasma on OPD performance**

**Fig. 1.** (a) shows the current density vs. applied bias, J(V), for the devices (shown in inset) without and with $O_2$ plasma treatment in blue and red, respectively. The current density is calculated using the active area measured by mapping the sensitivity of the diode as described elsewhere [22]. The $O_2$ plasma treatment induces a degradation of the light current density in the reverse bias regime and therefore of the diode sensitivity from 0.3 A/W to 0.14 A/W (at a light power of 0.3 W/m$^2$). A reduction of the current density can be due to the addition of deep traps leading to Shockley Read Hall (SRH) recombination. Deep traps due to disorder or impurities lead to recombination centers for electrons and holes, reducing the short circuit current in a solar cell [26]. In reverse bias regime, the recombination through SRH mechanisms is usually neglected as all carriers can be extracted [27]. However, if there is band bending at one interface, a low electric field is obtained in the layer and deep traps can be efficient in the reverse bias regime, leading to the recombination of generated electrons and holes. We note a one order of magnitude increase of the injection current at +2 V in the direct regime after plasma treatment. This evolution can be due to oxygen p-doping increasing the free carrier density. However, the dark current density in the reverse bias regime is not strongly influenced by the surface treatment.

The external quantum efficiency (EQE) measured at -2 V is given in **Fig. 1.(b)** for both diodes. This figure of merit highlights a strong impact of the surface treatment on the OPD performance. The EQE spectra exhibit two peaks around 440 and 660 nm reaching values of 62 and 65%, respectively, when no surface treatment is used. After $O_2$ plasma, the EQE is reduced to 30% at 440 nm and 46% at 660 nm, consistent with the degradation of the light current density observed in the J(V) characteristics. Measurements of the active layer absorption before and after $O_2$ plasma treatment (**Fig. S1.** Supplementary Information) show that this treatment does not induce any significant change in absorption in the visible part of the spectrum. Therefore, the changes in the EQE are not due to modifications in the absorption performances but in the carrier extraction efficiency.

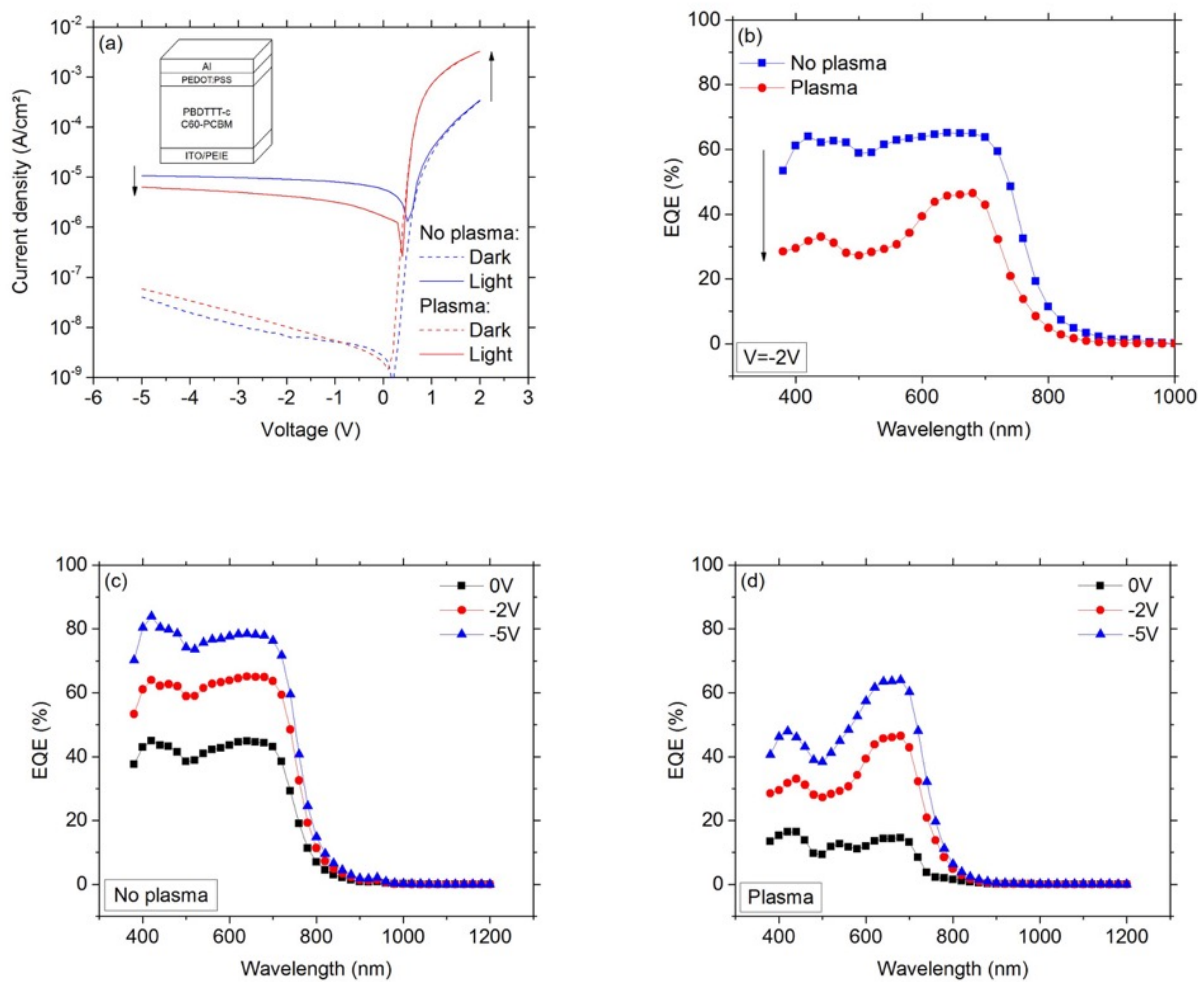

**Fig. 1.** Current density in the dark and under illumination (530 nm, 0.3 W/m$^2$) for the device without plasma treatment in blue and the device exposed to $O_2$ plasma in red (a). A schematic of the OPD structure is given as inset. External Quantum Efficiency (EQE) at -2 V (b) for the device without plasma treatment in blue and the device exposed to $O_2$ plasma in red. EQE at different biases for the device without plasma treatment (c) and the device exposed to $O_2$ plasma (d).

EQE measurements are carried out at different biases (0, -2 and -5 V) for both devices as shown in **Fig. 1.** (c) and (d). An increase in EQE with applied bias is observed for the device processed without plasma treatment and its shape remains independent of the electric field in the layer. The device exposed to $O_2$ plasma exhibits an EQE increase with applied bias as well but the electric field has an impact on the shape of the EQE, with the 660 nm peak undergoing a more pronounced increase than the 440 nm peak. This electric field dependence can be explained by charged traps in the bandgap, leading to band bending at one interface with the formation of a depletion zone. This hypothesis has already been developed by Wang et al. [25] with the active layer n-type doping induced by poly(ethyleneimine) (PEI), which is used to tune the cathode work-function.

To identify a potential p-doping of the active layer upon oxygen exposure, we carried out capacitance measurements to extract the hole density through a Mott-Schottky analysis. For a homogeneous doping density in the layer, the hole density p can be calculated using the following relation [28]:

$$p = -\frac{2}{q\varepsilon_0\varepsilon_r A^2 \frac{d(C^{-2})}{dV}} \quad (1)$$

where is the elementary charge, the vacuum permittivity, the relative permittivity of the blend and the active area of the diode. A relative permittivity of is extracted from C(V) measurements performed at 100 Hz in the reverse bias regime, when the whole layer is depleted and the geometric capacitance is reached. Knowing the diode area and the relative permittivity, the hole density is extracted for both types of devices. **Fig. 2.** (a) and (b) shows the Mott-Schottky plots for the devices processed without and with $O_2$ plasma, respectively. When no surface treatment is used, a hole density of is extracted around 0 V. Below -1 V, the Mott-Schottky plot saturates as the depletion width reaches the active layer thickness. However, Kirchartz et al. [29] have highlighted the limits of the Mott-Schottky extraction technique at low doping densities due to a violation of the depletion approximation. For a diode with a Schottky contact and a layer thickness around 300 nm, the Mott-Schottky technique is not appropriate for hole densities below . In our case of a 500 nm thick active layer, the density limit would be of the same order of magnitude. Therefore, it is difficult to conclude whether the extracted value of is reliable. The Mott-Schottky plot for the device exposed to $O_2$ plasma does not exhibit a linear regime suggesting that the doping concentration is not homogeneous in the layer. The hole density is then extracted at different bias ranges, leading to values around . The doping profile is reconstructed using the generalized Mott-Schottky law and is shown in **Fig. S2.** (Supplementary Information). Therefore we can conclude that an $O_2$ plasma treatment on the active layer before PEDOT:PSS deposition increases the hole density by at least one order of magnitude. Moreover, the depletion width follows a square root evolution with the applied bias (**Fig. S3.**, Supplementary Information), as expected for the formation of a depletion zone at the interface [30].

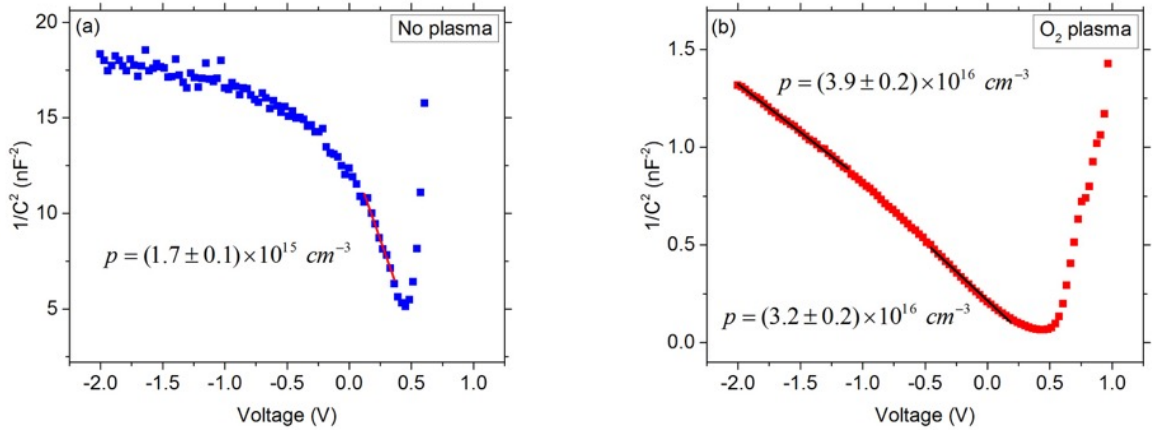

**Fig. 2.** Mott-Schottky plots obtained from C(V) measurements performed at 100 Hz for the device without plasma treatment (a) and the device exposed to $O_2$ plasma (b).

Electrical characterizations on photodetectors processed with and without $O_2$ plasma on the active layer highlight the impact of the surface treatment. The observed degradation can be explained by the formation of an acceptor state below the Fermi level in the bandgap. If the state is deep enough, it can lead to carrier recombination. If the acceptor state is filled with electrons, it induces doping and charged states leading to band bending and electric field dependency of the EQE. Finally, doping can explain the injection current increase in the direct regime. The following section is devoted to more detailed characterizations of p-doping and trap creation and their impacts on the OPD performance.

### 3.2. Conductivity and trap characterizations

To confirm the p-doping ability of the $O_2$ plasma treatment on the active layer, we use the transmission line method (TLM) to follow the evolution of the hole conductivity with plasma exposure time. A doping concentration increase should indeed induce a conductivity increase as both quantities are linked by  where  is the hole mobility [30]. Moreover, the structure depicted in **Fig. 3.** enables us to determine whether the $O_2$ plasma treatment is limited to the active layer surface or if the oxygen diffuses to the bulk of the layer. The contact resistance  and channel resistance  extracted from I(V) measurements for different channel lengths L and widths W (see **Fig. S4.** (a), (b) and (c), Supplementary Information) are given in Table 1. Three different treatment conditions are analyzed: without $O_2$ plasma, with a 60 s and with a 300 s treatment. The conductivity  is related to the series resistance  according to  with  the thickness of the electrodes (assuming that charge transport is mainly localized at the bottom interface). The conductivities for each treatment condition are summarized in Table 1. A blend exposed for 60 s to the $O_2$ plasma undergoes a conductivity increase by one order of magnitude. Increasing the exposure time to 300 s enhances the hole conductivity by an additional order of magnitude, reaching around at the bottom part of the layer. Since oxygen needs to diffuse over 470 nm to reach the TLM channel area, this analysis demonstrates that $O_2$ plasma affects the bulk as well as the surface. Moreover, the contact resistance decreases by two orders of magnitude after a 300 s of plasma exposure time, which is consistent with p-doping of the active layer, leading to the effective injection barrier lowering at the electrode/blend interface [31].

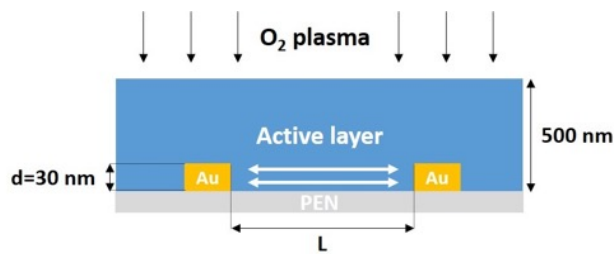

**Fig. 3.** Cross-section of the TLM structure showing the active layer and gold electrodes thicknesses. The variable channel length L and the surface exposed to the $O_2$ plasma treatment are also indicated.

**Table 1** Contact resistance, series resistance and conductivity extracted from TLM measurements for three different plasma treatment exposure times.

| $O_2$ plasma exposure time (s) | $R_C$ (Ω) | $R_S$ (Ω) | $\sigma_p$ (S/cm) |
|---|---|---|---|
| 0 s | $(8 \pm 4) \times 10^8$ | $(9 \pm 1) \times 10^{11}$ | $(3.9 \pm 0.4) \times 10^{-7}$ |
| 60 s | $(2.1 \pm 0.5) \times 10^7$ | $(7.1 \pm 0.1) \times 10^{10}$ | $(4.69 \pm 0.07) \times 10^{-6}$ |
| 300 s | $(6 \pm 1) \times 10^6$ | $(7.0 \pm 0.3) \times 10^9$ | $(4.76 \pm 0.02) \times 10^{-5}$ |

The evolution of EQE and light current density can be related to the formation of trap states in the bandgap after $O_2$ plasma treatment. This assumption is strengthened by the work of Knipp et al., which suggests the formation of an acceptor trap state 290 meV above the highest occupied molecular orbital (HOMO) of the organic semiconductor (pentacene in their case) when exposed to oxygen [17]. Moreover, trap states 53 and 100 meV above the blend HOMO have been measured for poly(3-hexylthiophene) (P3HT) with (6,6)-phenyl C61-butyric acid methyl ester ($C_{61}$PCBM) exposed to oxygen [32]. To probe the existence of trap states in the blend with and without $O_2$ plasma treatment, we use admittance spectroscopy. The technique consists in analyzing the capacitance C and conductance G as a function of frequency at different temperatures (from 200 to 300 K in this study). Defects in the bandgap impact both conductance and capacitance due to the response (trapping/de-trapping) of charges to the oscillating signal. Typically, a step in the C vs. ω and a peak in the G/ω vs. ω curves are observed at transition frequency . These features are temperature dependent and enable the extraction of parameters related to the traps. The activation energy of a trap can be extracted according to the following Arrhenius law [33]:

$$\omega_T = \nu_0 \exp\left(-\frac{E_A}{k_B T}\right) \quad (2)$$

with the attempt-to-escape frequency, the Boltzmann constant and the temperature.

To facilitate the extraction of the transition frequency , we plot with respect to the angular frequency with the DC conductance value, evaluated here at the lowest measured frequency (20 Hz). When trap states are probed by admittance spectroscopy, a peak appears in conductance spectra. The position of the peak depends on the activation energy and capture cross-section of the energy level. Frequency dependent conductance spectra are shown in **Fig. 4.** (a) and (b) without and with $O_2$ plasma treatment, respectively. To extract the activation energy using relation (2), we plot given by the position of the peak with respect to the inverse thermal energy . **Fig. 4.** (c) shows the Arrhenius plot for the OPD that was not exposed to $O_2$ plasma. The peak becomes visible only for temperatures higher than 280 K. An activation energy of approximately 290 meV is obtained, although the limited amount of data points makes it difficult to extract properly the parameters associated with the energy level. When exposed to $O_2$ plasma, the OPD exhibits sharp peaks and transition frequencies can be extracted from 240 to 300 K. The fitting of the Arrhenius plot in **Fig. 4.** (d) leads to an activation energy around 350 meV. The peak corresponding to the activation energy of 290 meV is no longer observed. The peak related to this trap state might be hidden below the large peak corresponding to the trap at 350 meV.

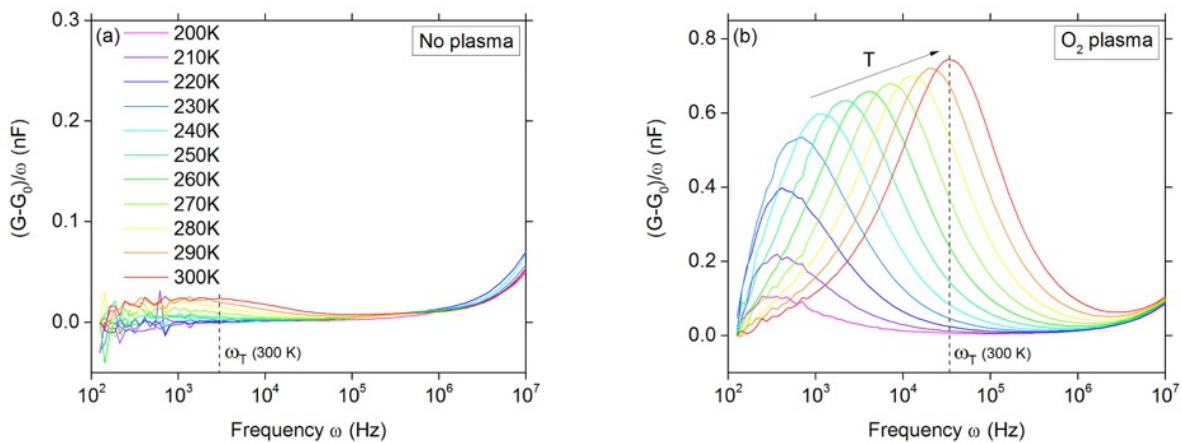

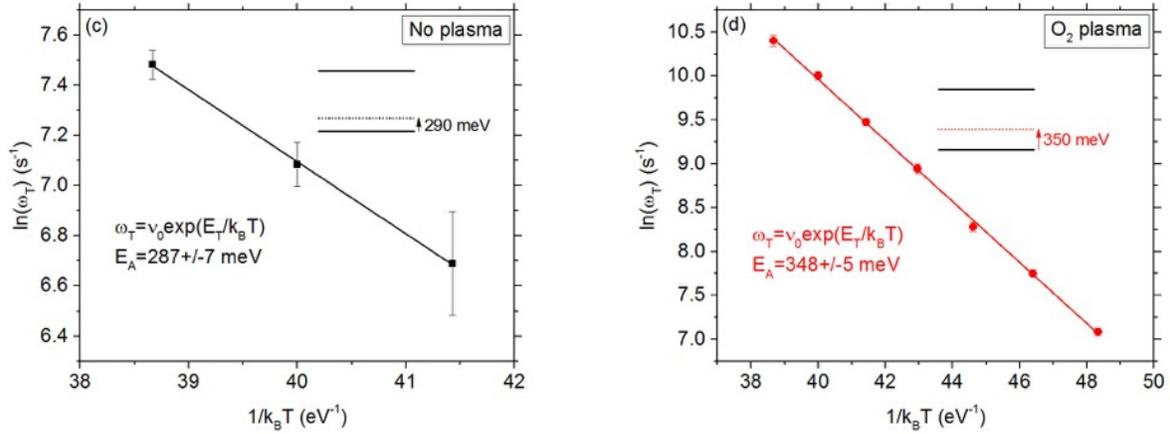

**Fig. 4.** versus angular frequency as a function of temperature at 0 V for the device without plasma (a) and with O$_2$ plasma exposure (b). Arrhenius plot derived from the admittance spectroscopy measurements on the samples without plasma (c) and with O$_2$ plasma exposure (d). The activation energies are extracted with a linear fit.

To determine whether the extracted activation energies are given with respect to the blend HOMO or lowest unoccupied molecular orbital (LUMO), we need to consider the electrodes work functions. A defect in the semiconductor bandgap is probed by admittance spectroscopy when the trap level is crossed by the Fermi level at equilibrium or by the quasi-Fermi levels when a bias is applied. In a diode structure, the quasi-Fermi levels are necessarily situated between the cathode and the anode work functions. As a consequence, states below the anode or above the cathode quasi-Fermi levels cannot be probed by this technique [34]. In that regard, The ITO/PEIE and PEDOT:PSS work functions are measured by Kelvin probe at 4.2 and 4.9 eV, respectively. The schematic in **Fig. 5.** illustrates the energy windows of the blend band gap that are presumably not probed by admittance spectroscopy (hatched area). Since the measured activation energies are both lower than the difference between the cathode work function and the LUMO of the blend, we can conclude that those energies are necessarily given with respect to the blend HOMO. A schematic of the energy level diagram with the corresponding trap state is given as inset in **Fig. 4.** (c) and (d).

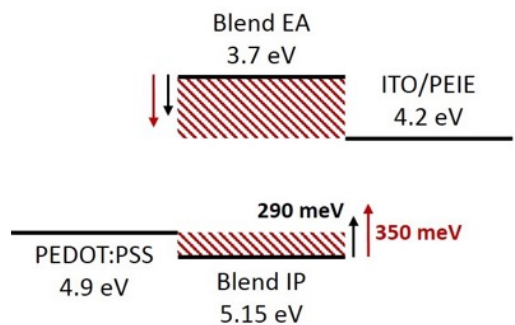

**Fig. 5.** Band diagram of the OPD showing the bandgap areas that cannot be probed by admittance spectroscopy (hatched area). The activation energies of 290 and 350 meV are illustrated with black and red arrows respectively from the blend HOMO and from the blend LUMO.

The concentration of traps can be calculated from the amplitude of the conductance peak at the transition frequency according to the relations and with the capacitance associated to the trap (trapped charges) and the variation of the depletion width with the oscillating signal [35]. When no plasma is used to deposit PEDOT:PSS, a density of is calculated for the trap level situated approximately 290 meV above the blend HOMO. For the device exposed to $O_2$ plasma, a trap concentration of is obtained for the level situated 350 meV above the blend HOMO. Admittance spectroscopy measurements are limited by the amount of carriers available to be trapped and detrapped. Since the values and carrier densities extracted from capacitance measurements (section 3.1) are close, the extracted values of might be underestimated.

Electrical characterizations carried out on devices processed with and without $O_2$ plasma treatment led to the hypothesis that $O_2$ plasma results in the p-doping of the active layer through oxygen diffusion in the bulk with the formation of an acceptor state situated below the Fermi level. We strengthened this hypothesis by further highlighting the blend p-doping in the bulk and the formation of a trap state 350 meV above the blend HOMO after plasma treatment. Supported by several studies presented in the literature [17,32,36], we suggest that the states observed by admittance spectroscopy are acceptor traps responsible for p-doping. This assumption will be further developed using TCAD simulations in the following section.

### 3.3. Device simulations

Numerical simulations were carried out using the commercial tools available in Silvaco environment. We performed finite element simulations using TCAD (Technology Computer Aided Design) in the software ATLAS enabling electro-optical simulations of our devices. All parameters used in the simulation are summarized in the Supplementary Information, Table S5.

As a first step, J(V) and EQE characteristics of the device processed without plasma treatment are used to optimize the fitting parameters and , conduction and valence density of states. In order to fit properly the light current density and EQE values, mid-gap states must be added to the structure. These effective traps account for charge carrier recombination centers that cannot be probed by admittance spectroscopy under the temperature/frequency range of our experiments (deeper states are observed at higher temperatures or lower frequencies in admittance spectroscopy). The energy level identified 290 meV above the blend HOMO through admittance spectroscopy measurements is added to the structure and leads to a good fit of the J(V) characteristics only if it is considered as donor state as acceptor states lead to an injection current too high in the direct regime. **Fig. 6.** (a) and (b) exhibits the J(V) and EQE characteristics of the device processed without plasma treatment with the best fits obtained for donor states situated 290 meV above the blend HOMO and mid-gap states.

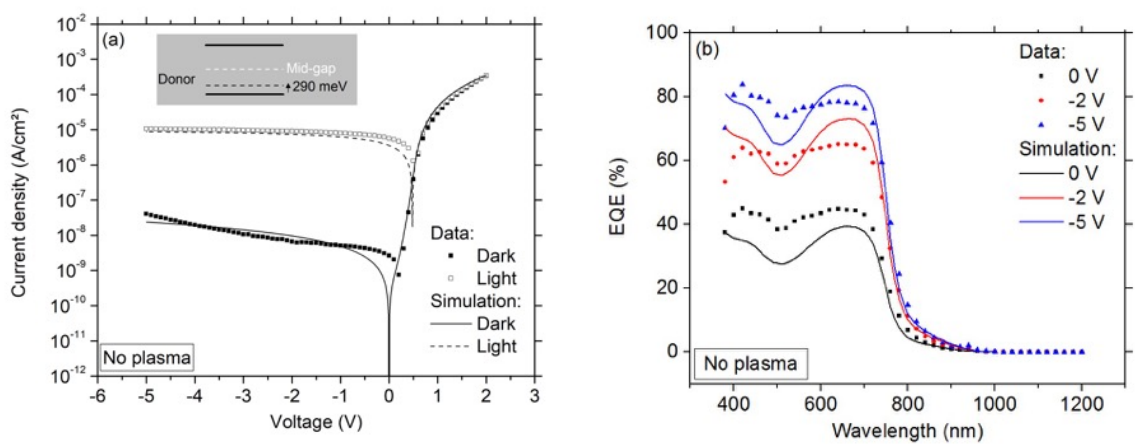

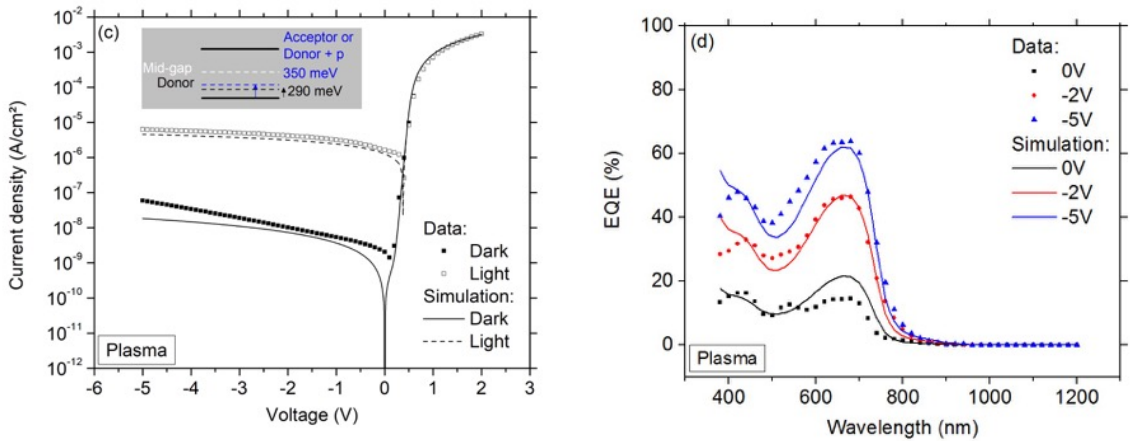

**Figure 6.** Current density in the dark and under illumination (530 nm, 0.3 W/m$^2$) (a) and EQE at 0, -2 and -5 V (b) for the device without plasma treatment. The simulations obtained with a donor trap level situated 290 meV above the blend HOMO are added to the graphs. Current density in the dark and under illumination (530 nm, 0.3 W/m$^2$) (c) and EQE at 0, -2 and -5 V (d) for the device processed with O$_2$ plasma treatment. The simulations obtained for the diode with plasma treatment are added to the graphs.

All parameters defined above remain unchanged for the simulation of the device processed with O$_2$ plasma treatment as the trap state situated 290 meV above the blend HOMO is considered intrinsic to the polymer. To fit the J(V) and EQE characteristics of this device, we add an additional energy level situated 350 meV above the blend HOMO, as measured by admittance spectroscopy. Two possibilities must be considered: the trap states are acceptor levels leading to p-doping or they correspond to defects induced donor levels accompanied by additional acceptor states. Acceptor states that would be situated too close to the blend HOMO cannot be probed by admittance spectroscopy in our experimental conditions (they should respond at higher frequencies or lower temperatures, but the signal is too weak <200K). To take these potential energy levels into account in the simulation, we define a hole density  corresponding to the quantity obtained by Mott-Schottky analysis. Both hypotheses lead to the exact same fits. Therefore, only one simulation is shown in **Fig. 6.** (c) and (d) along with the J(V) and EQE measurements for the device processed with O$_2$ plasma treatment. A good fit is obtained for an acceptor trap density or hole density of , consistent with experimental results ( in **Fig. 2.** And ). Although TCAD simulations cannot be used to distinguish between both hypotheses, it confirms that O$_2$ plasma treatment necessarily induce the active layer p-doping.

In order to better understand the impact of unintentional p-doping on the OPD electrical characteristics, we have simulated the band diagrams for both devices at -2 V. **Fig. 7.** (a) exhibits the band diagram corresponding to the diode processed without plasma treatment. A constant electric field can be observed in the active layer of this structure. When the blend is exposed to the O₂ plasma, a strong band bending is created at the cathode interface as shown in **Fig. 7.** (b).

To understand the impact of a potential band bending on the EQE evolution, we need to consider the absorption length in the active layer at different wavelengths. The absorption length corresponds to the depth of the layer required to absorb 63% of the incident flux and is defined by where corresponds to the wavelength of the incident photons and to the wavelength dependent extinction coefficient of the thin film. The active layer refractive index and extinction coefficient have been calculated from absorption measurements using the OptiChar module from OptiLayer Thin Film Software. The corresponding active layer absorption length with respect of the incident photons wavelength is given in Supplementary Information, **Fig. S6**. The absorption length varies between 80 and 360 nm depending on the incident wavelength. For low absorption lengths, the excitons are created in the high field area leading to an efficient dissociation and collection of charges. However, if the absorption length exceeds the depletion zone and reaches the flat band region, the probability of exciton recombination increases.

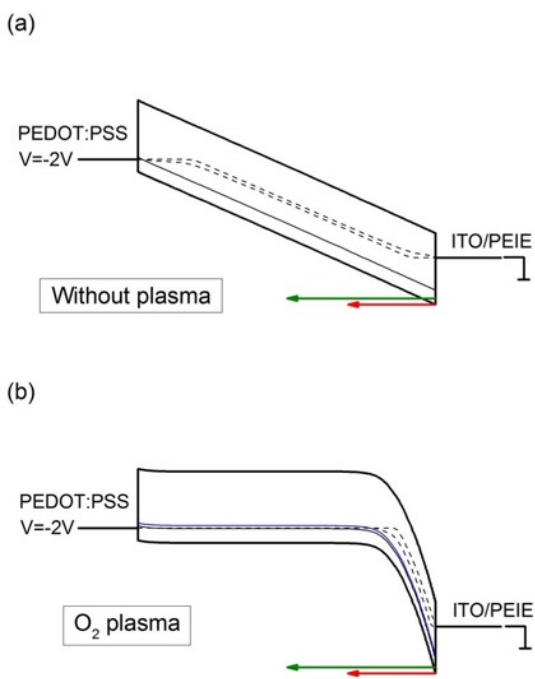

**Figure 7.** Band diagrams of the photodetectors without (a) and with (b) O₂ plasma treatment simulated at -2 V. The diagrams show the electrode work function, the active layer HOMO and LUMO (black solid line), the electrons and holes quasi-Fermi levels (dotted line) and the trap states. The trap level in black corresponds to the donor level situated 290 meV above the blend HOMO and the trap level in blue corresponds either to the acceptor level situated 350 meV above the blend HOMO or to the donor level at the same position with an additional hole density. The absorption length of 250 and 145 nm are illustrated with green and red arrows respectively.

The EQE spectra exhibit two peaks around 440 and 660 nm. The absorption lengths corresponding to these wavelengths are 250 and 145 nm respectively and are illustrated by green and red arrows on the band diagrams in **Fig. 7.** This schematic highlights the influence of the absorption length in a non-uniform electric field. The band bending at -2 V for the OPD processed with $O_2$ plasma extends over 140 nm below the ITO/PEIE electrode. With an absorption length of 145 nm at an incident wavelength of 660 nm, most of the light intensity is absorbed in the region of high electric field. The resulting collection of charges is efficient. However, for an incident wavelength of 440 nm, the absorption length of 250 nm is larger than the space charge region. The excitons generated by the photons reaching the flat band area have a higher probability to recombine leading to a low collection efficiency. The formation of a band bending therefore explains the modification of the EQE shape after plasma treatment.

4. **Conclusion**

In this work, we determine the impact of the widely used $O_2$ plasma treatment on the performance of an OPD device. Taking advantage of the lamination process adapted to PEDOT:PSS deposition [10], we compare the same structure processed with and without plasma treatment. Electrical characterizations highlight that the light sensitivity and EQE decrease while the injection current increases upon the $O_2$ plasma treatment. Moreover, we observe a change in the shape of the EQE spectra. The formation of acceptor states situated below the Fermi level could explain all the evolutions observed in the electrical characteristics. The increase in hole density and conductivity in the bulk confirm the p-doping effect of the $O_2$ plasma treatment through oxygen diffusion in the active layer. Using admittance spectroscopy, we identify the formation of a trap state situated approximately 350 meV above the blend HOMO after plasma treatment. However, whether this energy level is a disorder-induced donor state or the acceptor state responsible for p-doping is still unclear. Device simulations successfully reproduce J(V) and EQE characteristics for both diodes, strengthening the role of oxygen p-doping during plasma treatment. The simulated band diagrams allow us to understand the evolution of the EQE shape. The p-doping causes a strong band bending at the interface with ITO/PEIE and a low electric field closer to the PEDOT:PSS electrode. This band bending leads to an absorption length dependent photocurrent explaining a change in the shape of the EQE spectra. The study shows that the $O_2$ plasma surface treatment is detrimental to OPD performances. Unintentional oxygen doping must be avoided and intentional molecular doping needs to offer a limited tendency to diffuse.

## ACKNOWLEDGMENTS

We thank Sylvie Lepilliet (IEMN-CNRS) for her help on the admittance spectroscopy set-up and measurements. A.K. acknowledges funding from the National Science Foundation under grants DMR-1506097.

**SUPPLEMENTARY INFORMATION**

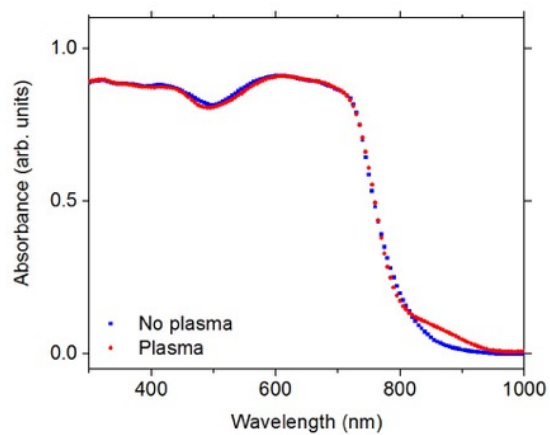

**Fig. S1.** Absorption spectra for PBDTTT-c:$C_{60}$-PCBM blend before (blue) and after (red) an $O_2$ plasma treatment of 60s.

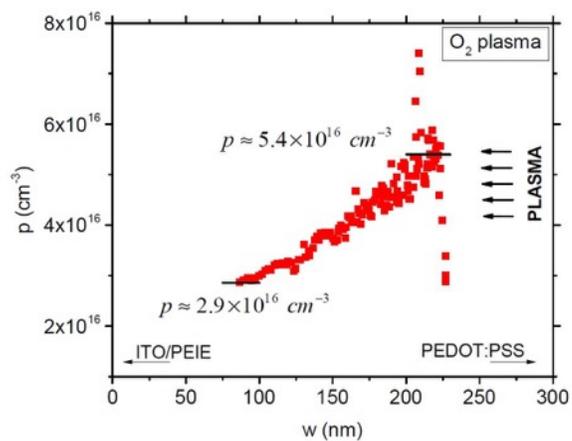

**Fig. S2.** Hole density profile in the active layer for the diode exposed to $O_2$ plasma.

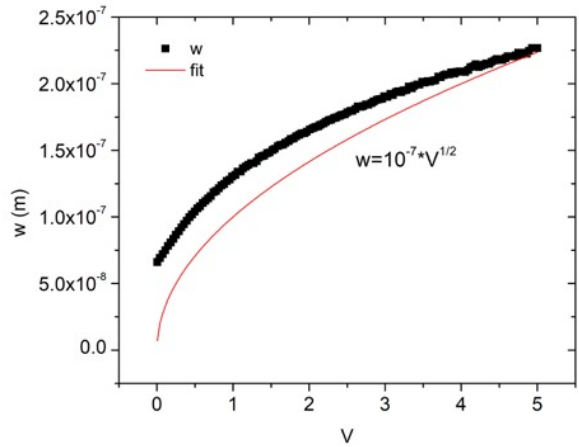

**Fig. S3.** Evolution of the depletion width w with the applied bias for the device processed with the $O_2$ plasma treatment. A fit following a square root law with the applied bias is given in red.

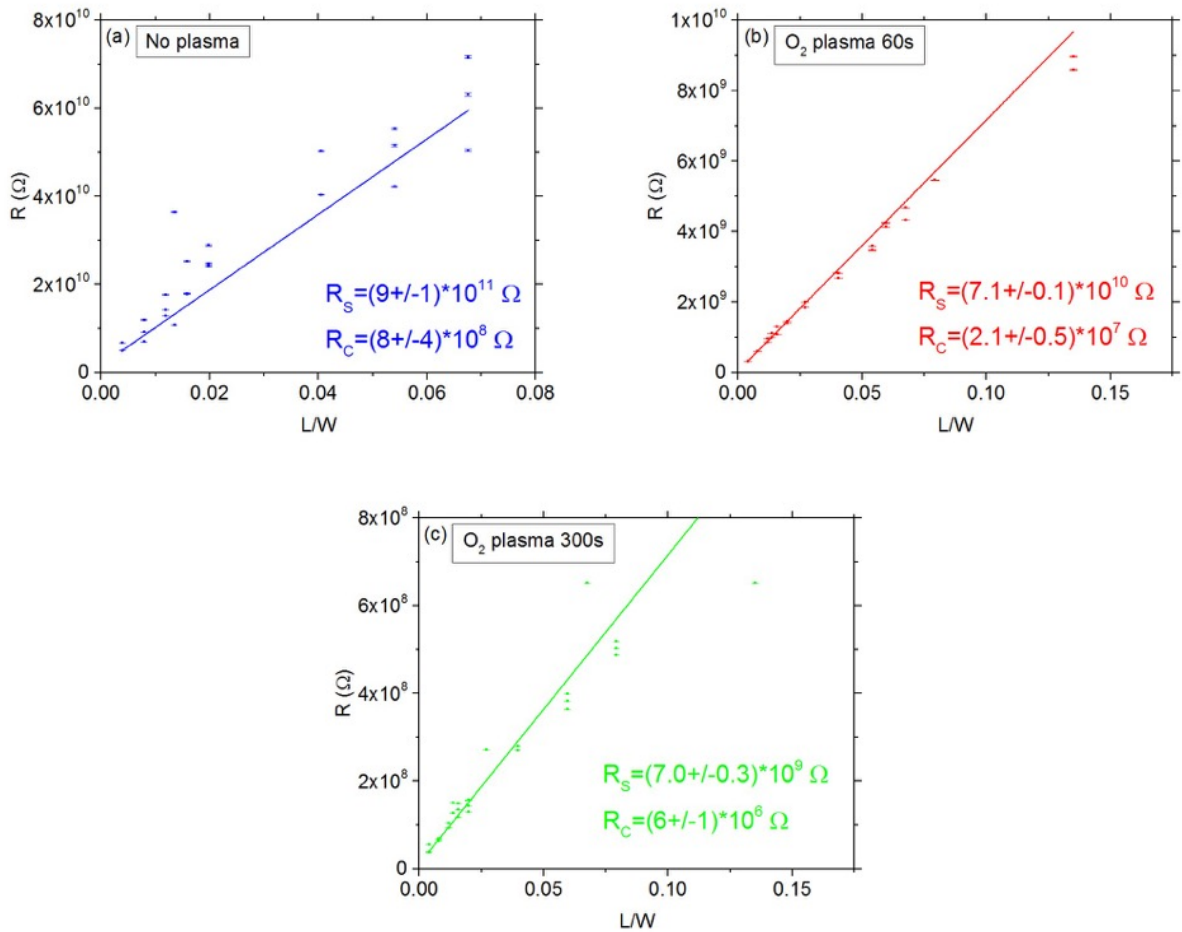

**Fig. S4.** Extraction of the contact and series resistances with respect to L/W ratio without plasma treatment (a), with a 60 s (b) and with a 300 s (c) $O_2$ plasma exposure time.

**Table S5.** Parameters used in the OPD simulation in Silvaco environment for both structures.

| Parameter | No plasma | With O$_2$ plasma | Source |
|---|---|---|---|
| Blend thickness | | | Contact profilometer |
| ITO/PEIE work-function | | | Kelvin probe |
| Blend electron affinity | | | Value from supplier (Sigma Aldrich) |
| Electrical bandgap | | | Calculated from HOMO measured by UPS |
| PEDOT:PSS work-function | | | Kelvin probe |
| Relative permittivity | | | C(V) measurements |
| Electron mobility | | | From literature [37] |
| Hole mobility | | | From literature [37] |
| Density of states and | | | Fitting parameter |
| Density of mid-gap states | | | Fitting parameter |
| Capture cross-section of mid-gap states () | | | Fitting parameter |
| Density of donor states 290 meV above the HOMO | | | Fitting parameter |
| Capture cross-section of the 290 meV trap | | | Fitting parameter |
| Density of acceptor or donor states 350 meV above the HOMO | | | Fitting parameter |
| Capture cross-section of 350 meV trap | | | Fitting parameter |
| Hole density if donor states | | | Fitting parameter |

I(V) and EQE characteristics of the device processed without oxygen plasma treatment are used to optimize the density of states and have a strong influence on the injected current density at low positive bias. Considering equivalent density of states in the HOMO and LUMO, a value of has been chosen. This value is close to the usual density of states around determined for organic semiconductors [38].

As we did not extract the electron and hole capture cross-sections of the trap states, these values are estimated with respect to the trap density by fitting the data. Moreover, the admittance spectroscopy does not lead to a good accuracy on the trap densities. As none of these variables are measured, we need to fix one of them to fit the second one. Therefore, there are multiple associations of capture cross-section and densities that could be suitable for the fitting. As a result, we cannot analyze the trap densities obtain by fitting the I(V) and EQE characteristics directly.

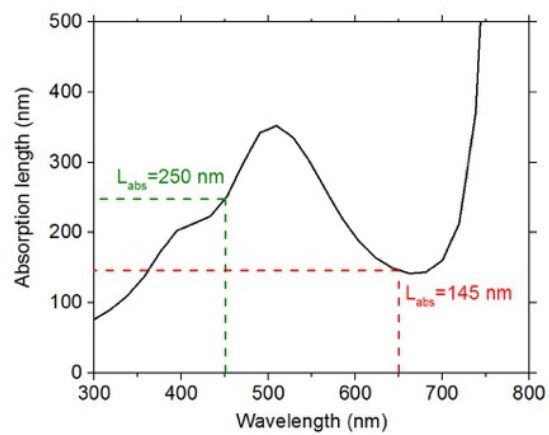

**Fig. S6.** Absorption length of PBDTTT-c:$C_{60}$-PCBM thin film with respect to the wavelength of the incident photons.